\documentclass{svjour2}
\journalname{arXiv}
\usepackage{amsmath}
\usepackage{amssymb}
\usepackage{amsfonts}

\begin{document}

\title{A lemma about molecular chaos}

\subtitle{Exact relations between many-particle correlations and
probability laws of diffusion in equilibrium ideal gas}

\author{Yuriy E. Kuzovlev}

\institute{Donetsk A.\,A.\,Galkin Institute for Physics and
Technology of NASU,
ul.\,R.Luxemburg 72, 83114 Donetsk, Ukraine\\
\email{kuzovlev@kinetic.ac.donetsk.ua}}

\date{August 19, 2009}
% The correct dates will be entered by Springer%

% Add name of the expert who has communicated your paper
%\communicated{name}

\maketitle

\begin{abstract}
Solutions to the BBGKY hierarchy of equations for molecular Brownian
particle in ideal gas are considered, and exact relations are derived
between probability distribution of path of the particle, its
derivatives in respect to gas density and irreducible many-particle
correlations of gas atoms with the path. It is shown that all the
correlations always give equally important contributions to evolution
of the path distribution, and therefore the exact statistical
mechanics theory does not reduce to classical kinetics even in the
low-density limit.
\end{abstract}

%\pagenumbering{arabic}
%\,\,\,
%\baselineskip 16 pt

\section{\,Introduction}
The idea of ``molecular chaos'' expressed by the Boltzmann's
well-known ``Sto{\ss}zahlansatz'' \cite{bol} is so much attractive
that Bogolyubov, after he formulated \cite{bog} the exact hierarchy
of evolution equations (now referred to as the
Bogolyubov-Born-Green-Kirkwood-Yvon, or BBGKY, hierarchy \cite{re})
for $\,s\,$-particle ($\,s=1,2,\dots\,$) distribution functions,
there and then truncated it at $\,s=2\,$ to examine possibilities of
substantiation of the Boltzmann equation for dilute gases. In fact,
however, until now the Boltzmann equation has no rigorous
substantiation based on the BBGKy hierarchy, and role of the
higher-order ($\,s>2\,$) distribution functions still is not properly
understood\,\footnote{ The frequently mentioned Lanford theorem
\cite{lan} about gas of hard spheres under the Boltzmann-Grad limit
concerns the so called ``hard-sphere BBGKY hierarchy'' (about it see
also e.g. \cite{re,vblls,cer}) which is not a true BBGKY hierarchy
since represents interactions of the spheres by invented terms like
the Boltzmann collision integrals (i.e. postulates what should be
proved, if any). Besides, the Lanford's result spans too short
evolution time intervals only.}\,.

In this communication I want to prove that no truncation of the BBGKY
hierarchy can assert a qualitatively correct statistical description of
gas evolution, even in case of arbitrary dilute gas. At that, in
order to simplify the proof and at once essentially strengthen it,
instead of the usual gas we will consider motion of a test particle
in thermodynamically equilibrium ideal gas whose molecules interact
with this particle only but not with each other (thus we concentrate
on situation least favorable for inter-particle correlations).

In Sec.2 and Sec.3 we formulate the BBGKY equations for this
system and then
rewrite them in terms of suitably defined ``cumulant distribution
functions'' responsible for irreducible $\,s$-particle
correlations between the test particle
and $\,s-1\,$ gas molecules ($\,s=1\,$ corresponds to the path
probability distribution of the test particle).
Next, show that these equations
imply exact relations between any $\,s$-particle
correlation (cumulant function) and derivative
of the previous $\,(s-1)$-particle one in respect
to the gas density. In Sec.4 we will make sure that, consequently, the
natural dimensionless
measure of any of the correlations keeps non-zero even under the
Boltzmann-Grad limit (hence, truncation
of the BBGKY hierarchy always is incorrect). Finally, some other
statistical properties of the correlations will be discussed.

\section{\,The BBGKY equations and cumulant distribution functions}
We want to consider thermal random motion of a test molecule (TM) in
thermodynamically equilibrium gas, with specified position $\,{\bf
R}(t)\,$ of the TM at some initial time moment $\,t=0\,$:\, $\,{\bf
R}(0)={\bf R}_0\,$.

Let $\,{\bf P}\,$ and $\,M\,$ denote momentum and mass of TM,
$\,m\,$, $\,{\bf r}_j\,$ and $\,{\bf p}_j\,$ ($\,j=1,2,...\,$) denote
masses, coordinates and momenta of other molecules,\, $\,\Phi({\bf
r})\,$ is (short-range repulsive) potential of interaction between
any of them and TM, and $\,n\,$ is gas density (mean concentration of
molecules). At arbitrary time $\,t\geq 0\,$, full statistical
description of this system is presented by the chain of
$\,(k+1)$-particle distribution functions ($\,k=0,1,2,...\,$):\,
$\,F_0(t,{\bf R},{\bf P}|\,{\bf R}_0\,;n\,)\,$\, which is normalized
(to unit) density of probability distribution of TM's variables, and
\,$\,F_k(t,{\bf R}, {\bf r}^{(k)},{\bf P},{\bf p}^{(k)}|\,{\bf
R}_0\,;n\,)\,$\, (where\, $\,{\bf r}^{(k)} =\{{\bf r}_1...\,{\bf
r}_k\,\}\,$, $\,{\bf p}^{(k)} =\{{\bf p}_1...\,{\bf p}_k\,\}\,$)
which are probability densities of finding TM at point $\,{\bf R}\,$
with momentum $\,{\bf P}\,$ and simultaneously finding out some
$\,k\,$ molecules at points $\,{\bf r}_j\,$ with momenta $\,{\bf
p}_j\,$. A rigorous definition of such distribution functions (DF)
was done in \cite{bog}. In respect to the coordinates $\,{\bf r}_j\,$
they are not normalized, but instead (as in \cite{bog}) obey the
conditions of decoupling of inter-particle correlations under spatial
separation of particles (in other words, DF satisfy a cluster
property with respect to spacial variables). Subject to the
symmetry of DF in respect to $\,x_j= \{{\bf r}_j,{\bf p}_j\}\,$\,
these conditions can be compactly written as follows:\,
$\,F_k\,\rightarrow\,F_{k-1}\,G_m({\bf p}_k)\,$\, at \,$\,{\bf
r}_k\rightarrow\infty\,$\,,\, where $\,G_m({\bf p})\,$ is the Maxwell
momentum distribution of a particle with mass $\,m\,$.

The enumerated DF satisfy a standard chain of equations \cite{bog}:
\begin{equation}
\frac {\partial F_k}{\partial t}=[\,H_{k},F_k\,]+n \,\frac {\partial
}{\partial {\bf P}}\int_{k+1}\!\!\Phi^{\,\prime}({\bf R}-{\bf
r}_{k+1}) \,F_{k+1}\,\,\,,\label{fn}
\end{equation}
with $\,k=0,1, ...\,$\, and along with obvious initial conditions
\begin{equation}
\begin{array}{c}
F_k|_{t=\,0}\,=\delta({\bf R}-{\bf R}_0)\, \exp{(-H_k/T\,)}= \label{ic}\\
= \delta({\bf R}-{\bf R}_0)\,G_M({\bf P}) \prod_{j\,=1}^k E({\bf
r}_j-{\bf R})\, G_m({\bf p}_j)\,\,,
\end{array}
\end{equation}
where\, $\,H_{k}\,$ is Hamiltonian of subsystem ``\,$k$ molecules +
TM\,'',\, $\,[...,...]\,$ means the Poisson brackets,\, $\,\int_k ...
=\int\int ...\,\,d{\bf r}_k\,d{\bf p}_k\,$\,,\,
$\,\Phi^{\,\prime}({\bf r}) =\nabla\Phi({\bf r})\,$\,, and $\,E({\bf
r})=\exp{[-\Phi({\bf r})/T\,]}\,$. Notice that TM can be considered
as a molecule of non-uniformly distributed impurity, and equations
(\ref{fn}) are identical to the equations of two-component gas
\cite{bog} in the limit of infinitely rare impurity, when the main
component is in spatially homogeneous and thermodynamically
equilibrium state.

Equations (\ref{fn}) together with (\ref{ic}) unambiguously determine
evolution of $\,F_0\,$ and eventually the probability distribution of
TM's displacement $\,{\bf R}-{\bf R}_0\,$. These equations will
become more clear if we make a linear change of DF $\,F_k\,$ by new
functions $\,V_k\,$ with the help of recurrent relations as follow:
\begin{equation}
\begin{array}{c}
F_0(t,{\bf R},{\bf P}| \,{\bf R}_0;n)\,=\,V_0(t,{\bf R},{\bf P}|\,
{\bf R}_0;n)\,\,\,,\\ F_1(t,{\bf R},{\bf r}_1,{\bf P},{\bf
p}_1|\,{\bf
R}_0;n)\,=\,\\
=\,V_0(t,{\bf R},{\bf P}|\,{\bf R}_0;n)\,f({\bf r}_1\!-{\bf R},{\bf
p}_1)\,+\, V_1(t,{\bf R},{\bf r}_1,{\bf P},{\bf p}_1|\,{\bf
R}_0;n)\,\,\,, \label{cf1}
\end{array}
\end{equation}
where\, $\,f({\bf r},{\bf p}) = E({\bf r})\,G_m({\bf p})\,$\,,
\[
\begin{array}{c}
F_2(t,{\bf R},{\bf r}^{(2)},{\bf P},{\bf p}^{(2)}|{\bf R}_0;n)\,
=V_0(t,{\bf R},{\bf P}|{\bf R}_0;n) \,f(\rho_1,{\bf
p}_1)\,f(\rho_2,{\bf p}_2)\,+\\+\,V_1(t,{\bf R},{\bf
r}_1,{\bf P},{\bf p}_1|{\bf R}_0;n) \,f(\rho_2,{\bf p}_2)
+ V_1(t,{\bf R},{\bf r}_2,{\bf P},{\bf p}_2|{\bf R}_0;n)
\,f(\rho_1\!,{\bf p}_1)\,+\\
+ \,V_2(t,{\bf R},{\bf r}^{(2)}, {\bf P},{\bf p}^{(2)}| \,{\bf
R}_0;n)\,\,\,,
\end{array}
\]
where\, $\,\rho_j\,\equiv\, {\bf r}_j\!-{\bf R}\,$,\, and so on.

Apparently, from viewpoint of the probability theory,\, $\, V_k\,$
represent a kind of cumulants, or semi-invariants, or cumulant
functions (CF). It is important to notice that if all these CF were
zeros then all conditional DF of gas, $\,F_k/F_0\,$,\, would be
independent on initial position $\,{\bf R}_0\,$ of TM and thus on its
displacement $\,{\bf R}-{\bf R}_0\,$. This fact makes visible very
interesting specificity of the CF\, $\,V_k\,$\,:\, they are not mere
correlations between instant dynamic states of TM and
$\,k\,$ gas molecules but simultaneously their irreducible
correlations with the total past TM's displacement.

\section{\,Evolution of many-particle correlations and their relation
to density derivatives of the path probability distribution}
In terms of the CF the BBGKY hierarchy acquires a more complicated
tridiagonal structure (we omit uninteresting algebraic details):
\begin{eqnarray}
\frac {\partial V_{k}}{\partial t}=[H_k,V_k]+n \,\frac {\partial
}{\partial {\bf P}}\int_{k+1}\!\! \Phi^{\,\prime}({\bf
R}-{\bf r}_{k+1})V_{k+1}+\nonumber\\
+\,T\sum_{j\,=1}^{k}\, \mathrm{P}_{kj}\,G_m({\bf p}_k)
\,E^{\prime}({\bf r}_k-{\bf R}) \left[\frac {{\bf P}}{MT}+\frac
{\partial }{\partial {\bf P}}\right ] V_{k-1}\,\,\,.\label{vn}
\end{eqnarray}
Here\, $\,E^{\prime}({\bf r})=\nabla E({\bf r})\,$,\, and\,
$\,\mathrm{P}_{kj}\,$\, symbolizes transposition of the pairs of
arguments $\,x_j\,$\, and $\,x_k\,$. On the other hand, initial
conditions (\ref{ic}) and the above-mentioned clustering
conditions \cite{bog} take very simple form:
\begin{equation}
\begin{array}{c}
V_0(0\,,{\bf R},{\bf P}|\,{\bf R}_0;\,n)\,=\delta({\bf R}-{\bf
R}_0) \,G_M({\bf P})\,\,,\\
V_{k}(0\,,{\bf R}, {\bf r}^{(k)},{\bf P},{\bf p}^{(k)}|\,{\bf
R}_0;n)=0\,\,,\label{icv}\\
V_k(t,{\bf R}, {\bf r}^{(k)},{\bf P},{\bf p}^{(k)}|{\bf
R}_0;n)\rightarrow0\,\,\,\,\,\,\texttt{at}\,\,\,\,{\bf
r}_j\rightarrow \infty
\end{array}
\end{equation}
($\,1\leq j\leq k\,$). Thus, as it should be with cumulants, CF
$\,V_k\,$ disappear under removal of already one of molecules.

From equations (\ref{vn}) as combined with the boundary and
initial conditions (\ref{icv}) it is clear that passage to the
limit in (\ref{icv}) realizes in an integrable way, so
that integrals\, $\,\widetilde{V}_{k}=\int_{k+1} V_{k+1}\,$\, are
finite. Let us consider them. By applying the operation $\,\int_k\,$
to equations (\ref{vn}) one easy obtains equations
\begin{eqnarray}
\frac {\partial \widetilde{V}_{k}}{\partial t}=[H_k,\widetilde{V}_k]+n
\,\frac {\partial }{\partial {\bf P}}\int_{k+1}\!\!
\Phi^{\,\prime}({\bf
R}-{\bf r}_{k+1})\,\widetilde{V}_{k+1}\,+\nonumber\\
+\,\,\frac {\partial }{\partial {\bf P}}\int_{k+1}\!\!
\Phi^{\,\prime}({\bf R}-{\bf r}_{k+1})\,V_{k+1}\,+\,\,\,\,\,\,\,\,\,
\,\,\,\,\, \,\,\,\,\,\, \label{vn1}\\
+\,T\sum_{j\,=1}^{k}\, \mathrm{P}_{kj} \,G_m({\bf p}_k) \
\,E^{\prime}({\bf r}_k-{\bf R}) \left[\frac {{\bf P}}{MT}+\frac
{\partial }{\partial {\bf P}}\right ] \widetilde{V}_{k-1}\nonumber
\end{eqnarray}
(with\, $\,k=0,1,...\,$). Because of (\ref{icv}) initial conditions
to these equations are zero:\, $\,\widetilde{V}_{k}(t=0)=0\,$\, at any
$\,k\,$.

Now, in addition to $\,\widetilde{V}_{k}\,$, let us consider
derivatives of CF in respect to the gas density,\, $\,V^{\prime}_{k}=
\partial V_{k}/\partial n\,$\,. It is easy to see that differentiation
of (\ref{vn}) in respect to $\,n\,$ yields equations for the
$\,V^{\prime}_{k}\,$ which exactly coincide with (\ref{vn1}) after
changing there $\,\widetilde{V}_{k}\,$ by $\,V^{\prime}_{k}\,$.
Besides, in view of (\ref{icv}), initial conditions to these
equations again all are zero:\, $\,V^{\prime}_{k}(t=0)=0\,$\, at any
$\,k\geq 0\,$. These observations strictly imply exact equalities\,
$\,V^{\prime}_{k}=\widetilde{V}_{k}\,$,\, or
\begin{eqnarray}
\frac {\partial }{\partial n}\,\, V_{k}(t,{\bf R}, {\bf r}^{(k)},{\bf
P},{\bf p}^{(k)}|\,{\bf R}_0;n)\,=\,\, \,\,\,\, \label{me}\\
=\,\int_{k+1} V_{k+1}(t,{\bf R}, {\bf r}^{(k+1)},{\bf P},{\bf
p}^{(k+1)}|\,{\bf R}_0;n)\,\,\,.\nonumber
\end{eqnarray}
This is main formal result of the present paper.
Notice that it
evidently confirms the assumed finiteness of integrals\,
$\,\widetilde{V}_{k}=\int_{k+1} V_{k+1}\,$\,.
Together with the CF's definition (\ref{cf1}),
it forms a lemma on the way to similar general theorems
of statistical kinetics of fluids.

\section{\,Discussion and resume}
The equalities (\ref{me}) contain the proof promised in Sec.1. Indeed,
they show, firstly, that all the many-particle
correlations between gas molecules and total path, or displacement,
of the test molecule (TM) really exist,
i.e. differ from zero. Secondly,
integral values of all the correlations, represented
in the natural dimensionless form,
have roughly one and
the same order of magnitude. Indeed, multiplying equalities
(\ref{me}) by $\,n^k\,$ and integrating them over
TM's momentum and all gas variables, we have
\begin{eqnarray}
n^k\, V_k(t,\Delta;n)\,\equiv \,n^k
\!\int_1\!\! ...\!\int_k \int\! V_{k}\,d{\bf P}\,=\,
n^k\,\frac {\partial^k V_0(t,\Delta;n)}{\partial n^k}
\,\sim\, c_k V_0(t,\Delta;n)\,\,,\,\,\nonumber
\end{eqnarray}
where\, $\,V_0(t,\Delta;n)  =\int V_0(t,{\bf R},{\bf P}|{\bf
R}_0;n)\,d{\bf P}\,$\, is just the probability density distribution
of the TM's displacement,\, $\,\Delta ={\bf R}-{\bf R}_0\,$\,,
and\, $\,c_k\,$ some numeric coefficients obviously
comparable with unit. Equivalently, unifying all CF
into one generating function, we can write
\begin{eqnarray}
V_0(t,\Delta;n)+
\sum_{k=1}^{\infty} \frac {u^k n^k}{k!}\, V_k(t,\Delta;n)=
V_0(t,\Delta;(1+u)\,n)\,\,\,. \label{gf}
\end{eqnarray}
Thus, distribution of total of the TM's random walk
``is made of its correlations'' with gas molecules
like the walk itself is made of collisions with them,
and hardly some of the correlations can be neglected
if we aim at completely correct analysis of the BBGKY equations.

We see also that characteristic volume occupied by the correlations
has an order of the specific volume:
$\,(\, |\int_1 ...\int_k \int V_{k}\,d{\bf P} \,|/V_0
\,)^{1/k} \,\sim\, n^{-1}\,$.\, In the Boltzmann-Grad limit,
$\,n\rightarrow \infty\,$,\, $\,r_0\rightarrow 0\,$,\,
$\,\pi r_0^2 n =1/\lambda =\,$const\,,\, it becomes
vanishingly small as
measured by the TM's mean free path $\,\lambda\,$.
But, nevertheless, it remains on order of effective volume
of the ``collision cylinder'',
$\,\sim \pi r_0^2\lambda\,$. This observation prompts that
$\,k$-particle correlations are concentrated just at those
particular subsets of $\,k$-particle phase space which correspond
to (real or virtual) collisions, and therefore their action
holds under the limit. The same is said by the equality (\ref{gf})
which also holds out. This becomes obvious if we take into account
that actually important parameter of the integrated CF under
the Boltzmann-Grad limit must be $\,\lambda\,$ instead of $\,n\,$
and rewrite (\ref{gf}) in the form
\begin{eqnarray}
W_0(t,\Delta;\lambda)+
\sum_{k=1}^{\infty} \frac {u^k }{k!}\, W_k(t,\Delta;\lambda)=
W_0(t,\Delta;\lambda/(1+u))\,\,\,, \nonumber
\end{eqnarray}
where\, $\,W_k(t,\Delta;\lambda)=\,\lim\, n^k\,V_k(t,\Delta;n)\,$.
Thus, nothing changes under the  Boltzmann-Grad limit.

It is necessary to underline that the correlations under our
attention are qualitatively different from correlations which appear
in standard approximations of the BBGKY hierarchy
and connect velocities of particles after collision
(see e.g. \cite{bal}). In our notations, a pair correlation of
such the kind would look nearly as\,
$\,V_1(t,{\bf R},{\bf r},{\bf P},{\bf p})=
F_1(t,{\bf R}^{\prime},{\bf r}^{\prime},
{\bf P}^{\prime},{\bf p}^{\prime})
-F_1(t,{\bf R},{\bf r},{\bf P},{\bf p})\,$,\,
where the primed variables describe the pre-collision state.
Clearly, because of the phase volume conservation during
collision, integration of this expression over
$\,\rho={\bf r}-{\bf R}\,$ and the momenta results in zero.
This observation shows that our correlations do
live in the configurational space and connect coordinates
and walks of particles (may be coexisting with statistical
independence of their velocities).
The role and statistical meaning of such correlations
were under investigation already in \cite{i1}
(and in principle even earlier in \cite{bk12}).
By their nature, they are attributes of spatially
inhomogeneous states and evolutions (evolution of
$\,V_0(t,\Delta;n)\,$ gives an example).

In view of all the aforesaid, we can suppose that the
Boltzmann-Lorentz equation \cite{re,vblls} and, moreover, the
Boltzmann equation in itself and its generalizations do not represent
a (low-density) limit of the exact statistical mechanical theory. The
classical kinetics is only its simplified probabilistic model (may be
excellent in one respects and caricature in others). Of course, in
the exact theory also molecular chaos does prevail. But here it is
much more rich, even in dilute gas, and does not keep within naive
probabilistic schemes.

\appendix

\section*{Appendix}
In the following, I append the Introduction and
Discussion and resume sections from the first variant of this
paper (titled ``A truth of molecular chaos'')
rejected by the JSP (without any review or explanations)
and by the CMP, on the grounds of that ``it is in contrast
with Lanford's theorem'' and it ``is not clear because assumptions,
heuristic ideas and non rigorous steps are not clearly distinguished''.

In order to conform to the latter remark, now I moved off all
the heuristics which was localized in the
Introduction and Discussion and resume sections
(other sections remain exactly as before).
Nevertheless, in my opinion, this heuristics is useful,
and readers can see it below. What is for the Lanford's theorem,
it was sharply commented in the above footnote. For more detail
comments see \cite{idg}) and also \cite{i1} where principles
of correct collisional (with the help of the Boltzamnn integrals)
description of spatially inhomogeneous evolution were formulated.
Here, I can add that all the results of the present
paper easy extend to hard-sphere interaction.

\subsection*{\,Introduction}
One of creators of the modern probability theory A.\,Kolmogorov
underscored \cite{kol} that in it ``\,{\it the concept of
independence of experiments fills most important place}\,'' and
``\,{\it correspondingly one of most important objectives of
philosophy of natural sciences}\,'' is ``\,{\it clearing-up and
refinement of those prerequisites under which one can treat given
phenomena as independent}\,''\,\footnote{The italics in quotes means
citations freely translated by me from Russian texts.}. Recall that
in the probability theory some random phenomena or quantities $\,A\,$
and $\,B\,$\, by definition are termed ``independent'' if their
probability distributions are independent, that is
$\,P(A,B)=P(A)\,P(B)\,$ \cite{kol}. But in natural sciences the
independence of phenomena $\,A\,$ and $\,B\,$\, is thought as absence
of cause-and-effect connections between them, that is absence of
their mutual influence. Does independence in this natural sense mean
independence in the sense of the probability theory?

Certainly not from the viewpoints of common sense and philosophy.
Merely because $\,A\,$ and $\,B\,$ which do not directly influence
one on another nevertheless both can be parts of the same random
event and thus turn out to be indirectly connected.

From the scientific point of view, it is natural to bring the same
question to the statistical mechanics. One of creators of modern
theory of dynamical systems and statistical mechanics N.\,Krylov
thoroughly analyzed it \cite{kr} and confirmed the negative answer.
He concluded that opinions that ``\,{\it phenomena which are
``manifestly independent'' should have independent probability
distributions}\,'',\, and the like,\, are nothing but ``\,{\it
prejudices}\,'' \cite{kr}.

Especially Krylov pointed \cite{kr} to the firmness of such
prejudices\,\footnote{That ``{\it are so habitual that even persons
who agreed with my argumentation usually automatically go back to
them when facing with a new question\,}''.}. Only it explains why the
molecular chaos hypothesis put forward by Boltzmann many years ago
\cite{bol} until now dominates kinetics although never was somehow
substantiated \cite{re,kac}. And why N.\,Bogolyubov, when he obtained
\cite{bog} an exact hierarchy of evolution equations for
$\,s\,$-particle distribution functions, straight away truncated his
equations at $\,s=2\,$ taking in mind their reduction to the
Boltzmann equation.

Undoubtedly, molecules of rarefied gas are independent in the natural
sense since almost surely have nothing common in the past.
Nevertheless they can be essentially dependent in the sense of the
probability theory. This is quite understandable \cite{i1}. As
colliding particles have no common history, there is no back reaction
of the gas to past collisions of any of them. Therefore arbitrary
long fluctuations in relative frequency of collisions are
allowable\,\footnote{``\,{\it ... relative frequencies of some
phenomenon along a given phase trajectory, generally speaking, in no
way are connected to probabilities}\,'' \cite{kr}.}. These
fluctuations just play the role of aforesaid random events producing
indirect statistical interdependencies between pairs (or groups) of
particles capable of being participators of one and the same
collision (or a cluster of successive collisions).

As the consequence, $\,P(A,B)  \neq P(A)\,P(B)\,$ where $\,P(A)\,$ is
probability of finding a molecule at (phase) point $\,A\,$ and
$\,P(A,B)\,$ is probability of finding simultaneously two molecules
at points $\,A\,$ and $\,B\,$. At that, relaxation of one-particle
distribution $\,P(A)\,$ is determined by the two-particle correlation
$\,P(A,B) - P(A)\,P(B)\,$. Relaxation of the latter just similarly
always (regardless of the gas rarefaction) is determined by
three-particle correlation. And so on up to infinity. Since during
time interval $\,t\,$ a molecule undergoes $\,\sim t/\tau \,$
collisions (with $\,\tau\,$ being characteristic free-flight time), a
correct description of gas evolution over this interval requires
taking into account $\,s$-particle correlations with at least\, $\,s
\lesssim t/\tau \,$. Hence, in practice, in contrast to the
conventional opinion, the whole hierarchy of equations deduced by
Bogolyubov \cite{bog} is necessary.

In \cite{i1} and other works\,\footnote{\label{p12}
Kuzovlev\,\,Yu.\,E.\,:\, On statistics and 1/f noise of Brownian
motion in Boltzmann-Grad gas and finite gas on torus.\, Part I.
Infinite gas,\, arXiv:\, cond-mat/0609515\,;\, Part II. Finite gas,\,
arXiv:\, cond-mat/0612325\,.\, See also \,Kuzovlev\,Yu.\,E.\,:\,
Kinetic theory beyond conventional approximations and 1/f-noise,\,
arXiv:\, cond-mat/9903350\,.} approximate solutions to this hierarchy
or, in other words, the Bogolyubov-Born-Green-Kirkwood-Yvon (BBGKY)
equations \cite{re} were suggested for the problem about random
wandering of a test molecule, and explanations were expounded why the
Boltzmann's hypothesis is wrong. The aim of the present communication
is to prove the statements of preceding paragraph without any
approximations. At that we will strengthen the proof and besides
simplify it due to replacing the usual gas by ideal gas whose
molecules interact with the test molecule only but not with each
other.

\subsection*{\,Discussion and resume}
The result (\ref{me}) contains the proof promised in Sec.1. Indeed,
equalities (\ref{me}) show, firstly, that all the many-particle
correlations between gas molecules and past displacement of test
molecule (TM) really exist, i.e. differ from zero. Secondly, all they
have roughly one and the same order of magnitude. For instance, if
comparing their integral values, due to (\ref{me}) we can write, in
natural dimensionless units,
\[
n^k\!\int_1\!\! ...\!\int_k \int\! V_{k}\,d{\bf P}\,=\,
n^kV_0^{(k)}(t,\Delta;n) \,\sim\, c_k V_0(t,\Delta;n)\,\,,
\]
where\, $\,V_0(t,\Delta;n)  =\int V_0(t,{\bf R},{\bf P}|{\bf
R}_0;n)\,d{\bf P}\,$\, is probability distribution of the TM's
displacement\, $\,\Delta ={\bf R}-{\bf R}_0\,$\,,
\,$\,V_0^{(k)}(t,\Delta;n)=\partial ^{\,k}V_0(t,\Delta;n)/\partial
n^k\,$\, are its derivatives in respect to gas density $\,n\,$\,,
and\, $\,c_k\,$ some numeric coefficients. Hence, all the
correlations are equally important, and none of them can be neglected
if we aim at knowledge about true statistics of TM's random walk.

For more details, let us suppose that $\,(s+1)$-particle correlation
is so insignificant that one can assign $\,V_s=0\,$ in (\ref{vn}). At
that, according to (\ref{vn})-(\ref{icv}), all higher-order
correlations also will be rejected. Then, obviously, according to
(\ref{me}), distribution $\,V_0(t,{\bf R},{\bf P}|{\bf R}_0;n)\,$ and
thus $\,V_0(t,\Delta;n)\,$ must depend on\, $\,n\,$\, definitely as
an $\,(s-1)$-order polynomial. But, from the other hand,
distribution\, $\,V_0\,$\, what follows from the truncated chain of
equations (\ref{vn}) certainly is absolutely non-polynomial function
of\, $\,n\,$\,. With taking into account that equalities (\ref{me})
do express exact properties of solutions to (\ref{vn})-(\ref{icv}) we
see that very deep contradiction is on hand.

This contradiction clearly prompts us that truncation of the BBGKY
hierarchy leads to qualitative losses in its solution.

Some possible losses already were characterized in \cite{i1} (and
principally even much earlier in \cite{bk12,bk3}) and in part filled
up in \cite{i1} as well as preprints$\,^4\,$ (it may be useful also
to see some of my recent works\,\footnote{\label{pre}
Kuzovlev\,\,Yu.\,E.\,:\, Virial expansion of molecular Brownian
motion versus tales of statistical independency,\, arXiv:\,
0802.\,0288\,;\, Thermodynamic restrictions on statistics of
molecular random walks,\, arXiv:\, 0803.0301\,;\, Molecular random
walks in a fluid and an invariance group of the Bogolyubov generating
functional equation,\, arXiv:\, 0804.2023\,;\,
Molecular random walks and invariance group of
the Bogolyubov equation (to appear).\, In these works (see also
references therein) another ways to similar and more strong results
were presented, in particular, an approach based on general
properties of the Liouville evolution operator and ``generalized
fluctuation-dissipation relations'' \cite{j12,p}.}\,). Therefore here
we confine ourselves (continuing 5-th paragraph
of Introduction) by remark
that cutting of the $\,(s+1)$-particle correlation means cutting of
$\,s$-th and higher statistical moments of fluctuations in relative
frequency of TM's collisions with gas molecules (in other words,
fluctuations in diffusivity of TM \cite{bk12}). At\, $\,s=2\,$ these
fluctuations are completely ignored, and such truncated equations
(\ref{vn}) yield a closed equation for $\,V_0(t,{\bf R},{\bf P}|{\bf
R}_0;n)\,$ which is equivalent to the Boltzmann-Lorentz equation
\cite{re}.

It is necessary to emphasize that above reasonings, as well as the
exact relations (\ref{me}), are indifferent to a degree of the gas
rarefaction. Consequently, one can state that the Boltzmann-Lorentz
equation (moreover, all the classical kinetics including the
Boltzmann equation and its generalizations) does not represent a
(low-density) limit of the exact statistical mechanical theory. The
conventional kinetics is only (more or less adequate or caricature)
probabilistic model of exact theory. Of course, in the latter also
molecular chaos does prevail. But here it is much more rich, even if
speaking about rarefied gas, and does not keep within naive
probabilistic logics.

\end{document}